\documentclass[ 
      ,draft            
  ]
  {aipproc}

\layoutstyle{6x9}
\usepackage{psfrag}
\newcommand{\be}{\begin{equation}}
\newcommand{\ee}{\end{equation}}
\newcommand{\bdm}{\begin{displaymath}}
\newcommand{\edm}{\end{displaymath}}
\newcommand{\bea}{\begin{eqnarray}}
\newcommand{\eea}{\end{eqnarray}}

\newcommand{\fs}{\; \; .}
\newcommand{\co}{\; \; ,}

\newcommand{\lvac}{\langle 0|\,}
\newcommand{\rvac}{\,|0\rangle}
\newcommand{\ind}{\scriptscriptstyle}

\begin{document}

\title{$\pi\pi$ scattering: theory is ahead of experiment}

\author{H.~Leutwyler}{
  address={Institute for
Theoretical Physics, University of Bern,\\ Sidlerstr. 5, CH-3012 Bern,
Switzerland\\
E-mail: leutwyler@itp.unibe.ch}
}

\begin{abstract}{I draw attention to a recent breakthrough in the field of low
  energy pion physics: the consequences of the hidden symmetry of the QCD
  Hamiltonian have successfully been incorporated in the general dispersive
  framework for the $\pi\pi$ scattering amplitude, which is due to Roy. The
  meagre experimental information about the imaginary parts at and above 0.8
  GeV suffices to unambiguously and accurately pin down the scattering
  amplitude at lower energies. The recent Brookhaven data on the reaction
  $K\rightarrow\pi\pi e\nu$ provide a significant test of the theory. They
  imply that the Gell-Mann-Oakes-Renner relation is approximately
  valid -- the bulk of the pion mass indeed originates in the quark
  condensate.  
\begin{center}
Excerpts from a talk given in honour of Arnulfo Zepeda\\ at the X Mexican
School of Particles and Fields, Playa del Carmen, México, 2002.\end{center}}
\end{abstract}

\maketitle

\section{Prelude}
Arnulfo Zepeda published a pioneering paper in 1972, together with M.A.B.~Beg
\cite{Beg Zepeda}. In this paper, it is shown that, in the chiral limit, the
charge radii of pions and nucleons contain infrared singularities. In
particular, the pion radius diverges logarithmically if the pion mass is sent
to zero,
\be\label{BZ} \langle r^2\rangle^\pi\hspace{-0.5em}
\rule[-0.3em]{0em}{0em}_{em}=\frac{1}{(4\pi  F_\pi)^2}\ln 
\frac{\Lambda^2}{M_\pi^2}+O(M_\pi^2)\fs\ee
The result contains what is called a chiral logarithm, whose coefficient is
determined by the pion decay constant, $F_\pi\simeq 92.4\,\mbox{GeV}$. In the
terminology of 
Chiral Perturbation Theory (ChPT), the formula gives the leading term in the
chiral expansion. In that framework, momentum independent quantities such as
the pion mass 
or the charge radii are expanded in powers of the quark masses $m_u$ and
$m_d$. As is well-known, the leading term in the chiral expansion of $M_\pi$ is
proportional to $\sqrt{m_u+m_d}$, while the expansion of $F_\pi$ starts with a
constant term. 
Since the scale $\Lambda$ of the logarithm in eq.~(\ref{BZ})
is independent of the quark masses, the formula states that the charge radius
of the pion diverges logarithmically if the quark masses $m_u,m_d$ are sent to
zero. The scale $\Lambda$ is related
to the effective coupling constant $\ell_6$:
\be\label{l6} \ln\frac{\Lambda^2}{M_\pi^2}=\bar{\ell}_6-1\fs\ee
In my talk on the {\it 
Electromagnetic form factor of the pion}, I discussed the progress made in
this field since 1972. The recent interest in this quantity originates in
the fact that a very accurate knowledge of the pion
form factor is needed if the uncertainty in 
the Standard Model prediction for the magnetic moment of the muon is to match
the fabulous experimental precision. 

The final state interaction theorem states that, in the
elastic region, the phase of the electromagnetic form factor of the pion
coincides with the P-wave phase 
shift of the scattering process\footnote{Bose
statistics forbids two neutral pions to occur in a configuration with
angular momentum $\ell = 1$, so that the reaction  $\pi^+\pi^-\rightarrow
\pi^0\pi^0$ is irrelevant here.} 
$\pi^+\pi^-\rightarrow \pi^+\pi^-$.  For this reason, 
$\pi\pi$ scattering plays an essential role in the analysis of the form
factor. In the following, I restrict myself to a discussion of the 
state of the art in $\pi\pi$ scattering. Concerning the
application of that knowledge to the form factor, I refer to \cite{Arkady}.

\section{Roy equations}
In ChPT, the scattering amplitude is expanded in powers of the momenta as well
as in the quark masses. The perturbation series has explicitly been worked out
to two loops, that is to next-to-next-to-leading order \cite{BCEGS}.
The masses $m_u$ and $m_d$ are very small, so that the
expansion in these variables converges rapidly. The expansion in powers of the
momenta may be viewed as an
expansion around the center of the Mandelstam triangle, that is a power 
series in $s-\frac{4}{3}M_\pi^2$ and $t-\frac{4}{3}M_\pi^2$. 
In view of the strong,
attractive interaction in the channel with $I = \ell=0$, the higher orders of
that expansion are sizeable already at threshold, $s=4M_\pi^2$,
$t=0$. The chiral representation does account for the threshold singularities
generated by two-pion states, but it accounts for resonances only indirectly,
through their contributions to the effective coupling constants.
Dispersive methods are needed to extend the
range of the chiral representation beyond the immediate vicinity of the
Mandelstam triangle. 

The method we are using to implement analyticity,
unitarity and crossing symmetry is by no means new. As shown by Roy more than
30 years ago  \cite{Roy}, these properties of the $\pi\pi$ scattering amplitude
subject the partial waves to a set of coupled integral equations. 
The equations involve two subtraction constants, which may be
identified with the two S--wave scattering lengths $a_0^0$, $a_0^2$. 
If these two constants are given, the Roy equations allow us to calculate the
scattering amplitude in terms of the imaginary parts above the "matching point"
$E_m=0.8\,\mbox{GeV}$. The available
experimental information suffices to evaluate the relevant dispersion
integrals, to within small uncertainties \cite{ACGL,Descotes}. In this
sense,  $a_0^0$, $a_0^2$ represent the essential parameters in low energy
$\pi\pi$ scattering. 

The Roy equations possess an entire
family of solutions, covering a rather broad spectrum of physically quite
different scattering amplitudes, because analyticity, unitarity and
crossing symmetry alone do not determine the subtraction constants, and the
experimental information about these is consistent with a broad range
of values. For this reason, previous Roy equation analyses invariably came up
with a family of representations for the scattering amplitude rather than a
specific one -- the main result established on the basis of this framework was
the resolution of the ambiguities inherent in phase shift analyses
\cite{Pennington Protopopescu,KLL}.  

The missing ingredient in the traditional Roy equation analysis is the fact
that the Hamiltonian of QCD is almost   
exactly invariant under the group SU(2)$_{\ind R}\times$SU(2)$_{\ind L}$ of
chiral rotations among the two lightest quark flavours.
As will be discussed in some detail in the next section, the values of the two
subtraction constants can be predicted very sharply on this basis. 
In effect, this turns the Roy equations into a framework that 
fully determines the low energy behaviour of the $\pi\pi$ scattering amplitude.
As an example, I mention that the P-wave scattering length and effective range
are predicted very accurately:
$a_1^1=0.0379(5) \,M_\pi^{-2}$ and $b_1^1=0.00567(13)\,M_\pi^{-4}$. The manner
in which the P-wave phase shift  
passes through 90$^\circ$ when the energy reaches the mass of the $\rho$
is specified within the same framework, as well as the behaviour of the
two S-waves.  The analysis reveals, for instance, that the iso-scalar S-wave
contains a pole on the second sheet and the position can be
calculated rather accurately: the pole occurs at
$E=M_\sigma-\frac{1}{2}\,i\,\Gamma_\sigma$,  with
$M_\sigma=470\pm 30\;\mbox{MeV}$ and $\Gamma_\sigma=590\pm 40\;\mbox{MeV}$
\cite{CGL}, etc.  

Many papers based on alternative approaches can be found in the 
literature.  Pad\'e approximants, for instance, continue to enjoy popularity
and the ancient idea that the $\sigma$ pole  
represents the main feature in the iso-scalar S-wave also found new adherents
recently. Crude models such as these may be of interest in connection with
other processes where the physics yet remains to be understood, but 
for the analysis of the $\pi\pi$ scattering amplitude, they cannot compete
with the systematic approach based on analyticity and chiral symmetry.
In view of the precision required in the determination of the pion form
factor, ad hoc models are of little use, because the theoretical uncertainties
associated with these are too large.

\section{Prediction for the $\pi\pi$ scattering lengths}
Goldstone bosons of zero momentum do not interact:  if the
quark masses $m_u,m_d$ are turned off, the S-wave scattering lengths
disappear, $a_0^0,a_0^2\rightarrow 0$. Like the mass of the pion, 
these quantities represent effects that arise from the breaking of the chiral
symmetry generated by the quark masses. In fact, as shown by Weinberg
\cite{Weinberg 1966}, $a_0^0$ and $a_0^2$ are proportional to the square of
the pion mass
\be\label{CA} a_0^0=\frac{7M_\pi^2}{32 \pi F_\pi^2}+O(M_\pi^4)\co\hspace{2em} 
a_0^2=-\frac{M_\pi^2}{16 \pi F_\pi^2}+O(M_\pi^4)\fs\ee
The corrections of order $M_\pi^4$ contain chiral logarithms. In the case of
$a_0^0$, the logarithm has an unusually large coefficient
\bdm a_0^0=\frac{7M_\pi^2}{32 \pi
  F_\pi^2}\left\{1+\frac{9}{2}\,\frac{M_\pi^2}{(4\pi
    F_\pi)^2}\,\ln\frac{\Lambda_0^2}{M_\pi^2} +O(M_\pi^4)\right\}\fs\edm
This is related to the fact that in the channel with $I=0$, current algebra
predicts a strong, attractive, final state interaction. The scale $\Lambda_0$
is determined by the coupling constants of the effective
Lagrangian of $O(p^4)$:
\bdm \frac{9}{2}\ln\frac{\Lambda_0^2}{M_\pi^2}=\frac{20}{21}\,
 \mbox{$\bar{\ell}_1$}+\frac{40}{21}\,
\mbox{$\bar{\ell}_2$}-\frac{5}{14}\,
\mbox{$\bar{\ell}_3$}+
2\,\mbox{$\bar{\ell}_4$}+\frac{5}{2}\fs\edm 
The same coupling constants also determine the first order correction in the
low energy theorem for $a_0^2$.  

The couplings $\bar{\ell}_1$ and $\bar{\ell}_2$ control the momentum
dependence of the scattering amplitude at first non-leading order. Using the
Roy equations, these constants can be determined very accurately \cite{CGL}.
The terms $\bar{\ell}_3$ and $\bar{\ell}_4$, on the other hand, 
describe the dependence of the scattering amplitude on the quark masses --
since these cannot be varied experimentally,  $\bar{\ell}_3$ and
$\bar{\ell}_4$ cannot be
determined on the basis of $\pi\pi$ phenomenology. 
The constant $\bar{\ell}_3$ 
specifies the correction in the Gell-Mann-Oakes-Renner relation  \cite{GMOR},
\be\label{Mpi} M_\pi^2= M^2\left\{1-\frac{1}{2}\,\frac{M^2}{(4\,\pi
      F)^2}\;\bar{\ell}_3+O(M^4)\right\}\fs\ee
Here $M^2$ stands for the term linear in the quark masses,
\be\label{GMOR} M^2=(m_u+m_d)\,|\lvac \bar{u} u\rvac|\,\frac{1}{F^2}\ee
($F$ and $\lvac \bar{u}u\rvac$  are the values of the pion decay constant and
the quark condensate in the chiral limit, respectively). 
The coupling constant $\bar{\ell}_4$ occurs in the analogous expansion for 
$F_\pi$, 
\be\label{Fpi}
F_\pi =  F\left\{1+\frac{M^2}{(4\,\pi F)^2}\;\bar{\ell}_4
+O(M^4)\right\}\fs\ee
A low energy theorem relates it to the scalar radius
of the pion \cite{GL},
\be\label{radius} 
\langle r^2\rangle\!\rule[-0.3em]{0em}{0em}_s=\frac{6}{(4\pi F_\pi)^2}\left\{
\bar{\ell}_4-\frac{13}{12}+O(M^2)\right\}\co\ee
a formula that is very similar to one of Beg and Zepeda in eq.~(\ref{BZ}). 
The dispersive analysis of the scalar pion form factor in
ref.~\cite{CGL} leads to
\be\label{nradius} \langle r^2\rangle\!\rule[-0.3em]{0em}{0em}_s=0.61\pm0.04\;\mbox{fm}^2
\fs\ee
The constants $\bar{\ell}_1,\ldots \bar{\ell}_4$ depend logarithmically on the
quark masses:
\bdm \bar{\ell}_i=\ln\frac{\Lambda_i^2}{M^2}\co\hspace{2em}i=1,\ldots,4\edm
In this notation, the above value of the scalar radius amounts to 
\be\label{Lambda4} \Lambda_4 =1.26 \pm 0.14 \;\mbox{GeV}\fs\ee
Unfortunately, the constant $\bar{\ell}_3\leftrightarrow\Lambda_3$
is not known with comparable precision. The crude estimate for
$\bar{\ell}_3$ given in ref.~\cite{GL} corresponds to
\be\label{Lambda3} 0.2\;\mbox{GeV}<\Lambda_3< 2\;\mbox{GeV}\fs\ee
\setcounter{figure}{1}
\begin{figure}[thb]
\psfrag{a00}{\raisebox{-1em}{$a_0^0$}}
\psfrag{a20}{$a_0^2$}

\vspace{-3em}
\hspace{-1em}\includegraphics[width=9cm,angle=-90]{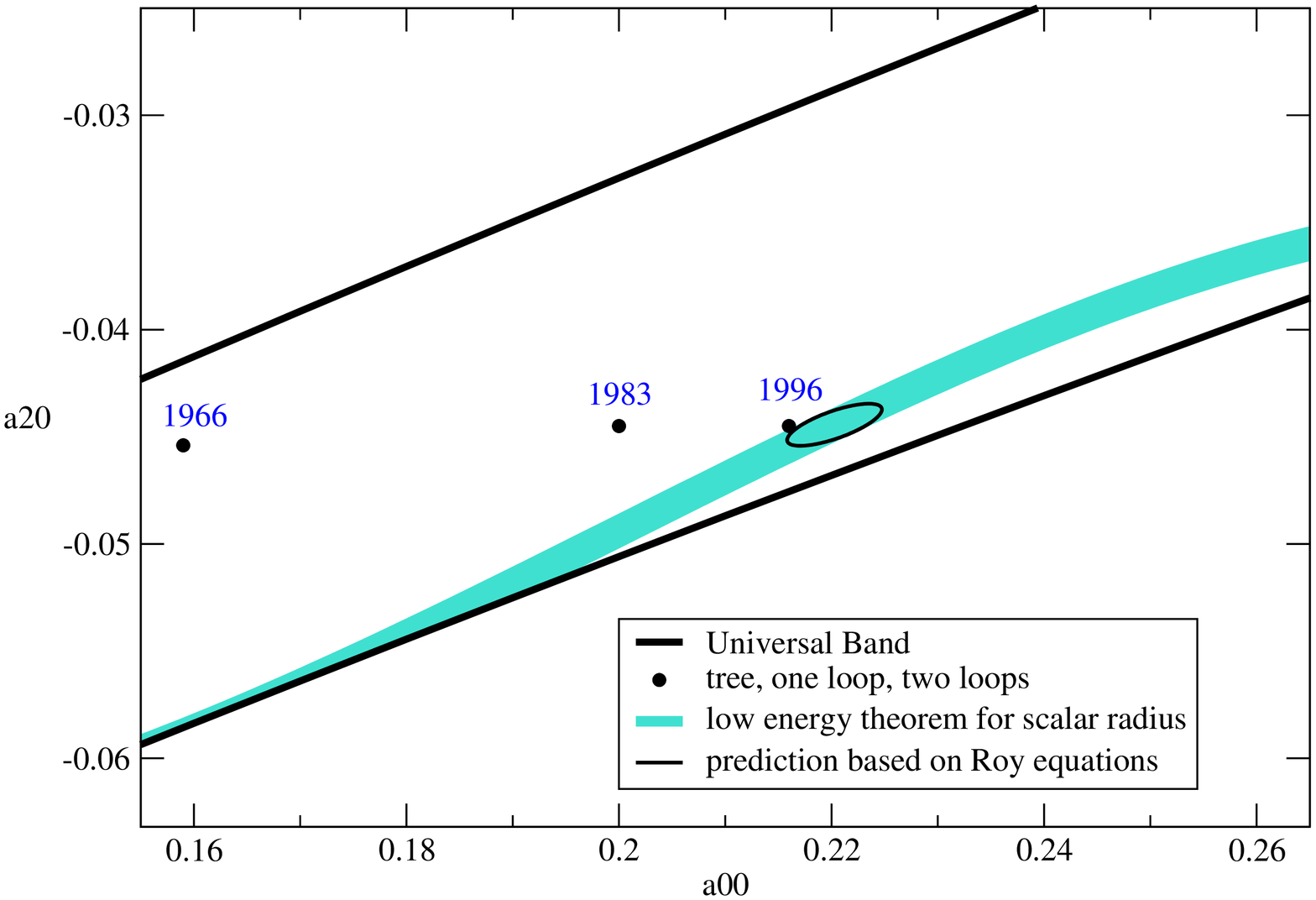}

\caption{Prediction for the S-wave $\pi\pi$ scattering lengths, taken from
  ref.~\cite{CGL}.} 
\end{figure}
It turns out, however, that the contributions from $\bar{\ell}_3$ are very
small, so that 
the uncertainty in $\Lambda_3$ does not strongly affect the
predictions for the scattering lengths. The result obtained in
ref.~\cite{CGL} reads
\be\label{a0a2} a_0^0=0.220\pm 0.005\,,\hspace{4em}a_0^2=-0.0444\pm
0.0010\,. \ee 
The analysis of the $K_{e_4}$ form factors reported in ref.~\cite{ABT} led to 
very similar results. Since that determination invokes an
expansion not only in $m_u$ and $m_d$, but also in $m_s$, it does not quite
reach the precision of the method underlying eq.~(\ref{a0a2}). 

In fig.~2, the
"Universal Band" shows the region in 
the $(a^0_0,a_0^2)$-plane where the Roy equations at all admit solutions. The
dot at the left represents the leading order result in eq.~(\ref{CA}), while
the small ellipse corresponds to the values in eq.~(\ref{a0a2}).

\section{Experimental test}\label{experimental test}
Stern and collaborators \cite{KMSF}
pointed out that "Standard ChPT" relies on a hypothesis that calls for 
experimental test.
Such a test has now been performed and I wish to briefly describe this
development. 

The hypothesis in question is the assumption that the quark
condensate represents the leading order parameter of the spontaneously broken 
chiral symmetry. More specifically, the standard analysis assumes that
the term linear in the quark masses dominates the expansion of
$M_\pi^2$. According to the
Gell-Mann-Oakes-Renner relation (\ref{GMOR}), this term is proportional to the
quark condensate, which 
in QCD represents the order parameter of lowest dimension. The dynamics of the
ground state is not well understood. The question raised by Stern et al.~is
whether, for one reason or the other,
the quark condensate might turn out to be small, so that the 
Gell-Mann-Oakes-Renner formula would fail -- the "correction" might be 
comparable to or even larger
than the algebraically leading term. 

According to eq.~(\ref{Mpi}), the
correction is determined by the effective coupling constant $\bar{\ell}_3$.
The estimate (\ref{Lambda3}) implies that the correction amounts to at most
4\% of the leading term, but this does not answer the question, because that
estimate is based on the standard framework, where 
$\lvac\bar{u}u\rvac$ is assumed to represent the leading order parameter. 
If that estimate is discarded and $\bar{\ell}_3$ is treated as a free
parameter ("Generalized ChPT"), the scattering lengths cannot be predicted 
individually, but the low energy theorem (\ref{radius})
implies that -- up to corrections of next-to-next-to leading order -- 
the combination $2a_0^0-5a_0^2$ is determined by the scalar
radius:
\bdm 2 a_0^0-5a_0^2=\frac{3M_\pi^2}{4\,\pi F_\pi^2}\left\{1+
\frac{M_\pi^2\langle r^2\rangle\!\rule[-0.3em]{0em}{0em}_s}{3}+\frac{41
  M_\pi^2}{192\,\pi^2F_\pi^2}+O(M_\pi^4)\right\}\fs\edm
The resulting correlation between $a_0^0$ and $a_0^2$ 
is shown as a narrow strip in fig.~2 (the strip
is slightly curved because the figure accounts for the corrections of
next-to-next-to leading order).  

In view of the correlation between $a_0^0$ and $a_0^2$, the data taken by the
E865-collaboration at Brookhaven \cite{Brookhaven} allow a significant 
test of the Gell-Mann-Oakes-Renner relation. The final state interaction
theorem implies that the phase of the form factors relevant for the
decay $K^+\rightarrow \pi^+\pi^- e^+\bar{\nu}_e$ is determined by the elastic
$\pi\pi$ scattering amplitude. Conversely, the phase difference
$\delta_0^0-\delta_1^1$ can be measured in this decay. The analysis of the
$4\cdot 10^5$ events of this type collected by E865 leads to the 
round data points
in fig.~3, taken from ref.~\cite{PRLCGL} (the triangles represent the
$K_{e_4}$ data collected in the seventies of the last century). 

\vspace{0.6em}
\begin{figure}[thb]
\centering
\includegraphics[width=9 cm]{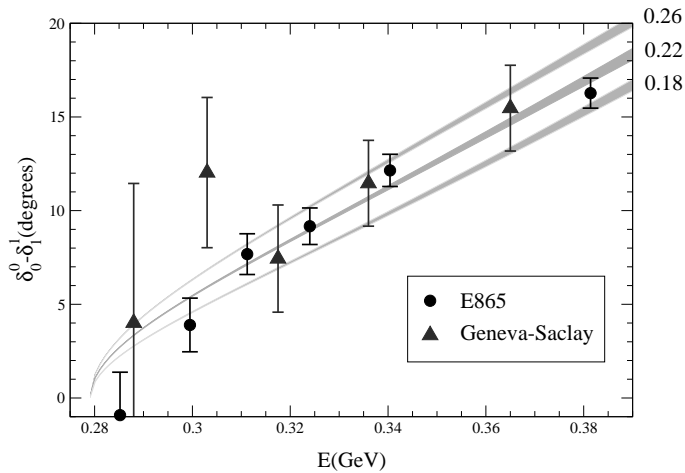}

\caption{Interpretation of the data on the phase difference
  $\delta^1_1-\delta^0_0$ in Generalized ChPT.}
\end{figure}
The three bands show the result obtained for
$a_0^0=0.18,0.22, 0.26$, respectively. The width of the bands corresponds to
the uncertainty in the prediction. A fit of the data that exploits the
correlation between $a_0^0$ and $a_0^2$  
yields 
\bdm a_0^0=0.216\pm 0.013\, (\mbox{stat})\pm
0.004\,(\mbox{syst})\pm0.005 \,(\mbox{th})\;\; \mbox{\cite{Brookhaven}},\edm
where the third error bar accounts for the theoretical uncertainties. The
result thus beautifully confirms the prediction of ChPT in eq.~(\ref{a0a2}). 
The agreement implies that more than 94\% of the pion
mass originate in the quark condensate, thus confirming that the
Gell-Mann-Oakes-Renner relation is approximately valid \cite{PRLCGL}. 
May Generalized ChPT rest in peace.

\section{Conclusion}In view of the fact that the pions are by far the lightest
hadrons, they play a prominent role in low energy physics. Chiral Perturbation
Theory offers a systematic approach for analyzing their properties. For the
$\pi\pi$ scattering amplitude, the perturbation series has been worked out to
two loops, but the resulting representation has two serious limitations: 
(1) it yields a decent approximation only at very low energies and (2) it
contains quite a few effective coupling constants, which need to be determined
in order to fully specify the scattering amplitude. 

In the course of the last two years, a breakthrough was achieved
here: we now have an accurate representation of the scattering amplitude
that (1) is valid in a significantly wider energy range than the chiral
representation and (2) 
does not contain any free parameters. In the vicinity of the center of the
Mandelstam triangle, where the chiral perturbation series is rapidly
convergent, the new representation agrees with the two-loop result of
ChPT. 

The new representation exploits the fact that the dependence of the $\pi\pi$
scattering amplitude on the momenta is very 
strongly constrained by general kinematics (unitarity, analyticity,
crossing). Although the Roy equations do not fully exhaust these constraints,
they do yield a very suitable framework for the low energy analysis. In this
framework, the S-wave scattering lengths $a_0^0$ and $a_0^2$ represent the
essential low energy parameters. They enter as 
subtraction constants of the partial wave dispersion relations -- once these
are known, the available 
experimental information suffices to accurately calculate the partial waves
below 0.8 GeV. In this context, ChPT is
needed exclusively to determine the two subtraction constants. 

Indeed, ChPT does yield very sharp predictions for
$a_0^0$ and $a_0^2$ \cite{CGL}. These predictions are based on the standard
framework, where it is assumed that the quark condensate is the leading order
parameter. That hypothesis has now been confirmed by the E865 data on the decay
$K\rightarrow \pi\pi e\nu$. More precise
data on this decay are forthcoming from experiment NA48/2 at CERN.  

A beautiful and qualitatively quite different experiment is also under
way at CERN \cite{Dirac}. There, charged pions are produced in abundance.
Occasionally, a pair of these binds to a $\pi^+\pi^-$ atom, "pionium". The
atoms almost instantly  decay
through the strong transition $\pi^+\pi^-\rightarrow\pi^0\pi^0$. Since the
momentum of the pions circulating in pionium is very small, of order $\alpha
M_\pi$, the transition amplitude is determined by the S-wave scattering
lengths: the decay rate is proportional to $(a_0^0-a_0^2)^2$. The
interplay of the electromagnetic and strong interactions in bound state and
decay is now very well understood \cite{Gasser et al}. A measurement of the
pionium lifetime at the planned accuracy of 10\% thus yields a measurement of
$a_0^0-a_0^2$ at the 5\% level, thereby providing a very sensitive test of the
prediction. DIRAC is a fabulous laboratory for low energy pion physics --
it would be most deplorable if this beautiful project were aborted for
financial reasons before its physics potential is tapped. In particular, 
pionium level splittings would offer a clean and direct measurement of the
second subtraction constant. Data on $\pi K$
atoms would also be very valuable, as they would allow us to explore the role
played by the strange quarks in the QCD vacuum. 
\begin{theacknowledgments}
I thank the organizers of the school for the invitation and
for their kind hospitality at Playa del Carmen and I am indebted to the
Humboldt Foundation for support.
\end{theacknowledgments}

\bibliographystyle{aipproc}   


\end{document}